\documentclass[showpacs,nofootinbib,prd,twocolumn]{revtex4}
\usepackage{graphicx}
\usepackage{amsfonts}
\usepackage{amssymb}
\usepackage{amsmath}

\begin{document}

\title{Application of the Campbell-Magaard
theorem \\ to higher-dimensional physics}
\author{Sanjeev S.~Seahra}
\email{ssseahra@uwaterloo.ca}
\author{Paul S.~Wesson}
\email{wesson@astro.uwaterloo.ca}
\affiliation{Department of
Physics, University of Waterloo\\ Waterloo, Ontario, N2L 3G1,
Canada}
\date{February 5, 2003}

\setlength\arraycolsep{2pt}
\newcommand*{\di}{\partial}
\newcommand*{\al}{\alpha}

\renewcommand{\textfraction}{0.15}
\renewcommand{\topfraction}{0.6}

\begin{abstract}

Stated succinctly, the original version of the Campbell-Magaard
theorem says that it is always possible to locally embed any
solution of 4-dimensional general relativity in a 5-dimensional
Ricci-flat manifold. We discuss the proof of this theorem (and
its variants) in $n$ dimensions, and its application to current
theories that postulate that our universe is a 4-dimensional
hypersurface $\Sigma_0$ within a 5-dimensional manifold, such as
Space-Time-Matter (STM) theory and the Randall \& Sundrum (RS)
braneworld scenario.  In particular, we determine whether or not
arbitrary spacetimes may be embedded in such theories, and
demonstrate how these seemingly disparate models are
interconnected. Special attention is given to the motion of test
observers in 5 dimensions, and the circumstances under which they
are confined to $\Sigma_0$. For each 5-dimensional scenario
considered, the requirement that observers be confined to the
embedded spacetime places restrictions on the 4-geometry.  For
example, we find that observers in the thin braneworld scenario
can be localized around the brane if its total stress-energy
tensor obeys the 5-dimensional strong energy condition. As a
concrete example of some of our technical results, we discuss a
$\mathbb{Z}_2$ symmetric embedding of the standard
radiation-dominated cosmology in a 5-dimensional vacuum.

\end{abstract}

\pacs{04.50.+h, 04.20.Cv}

\maketitle

\section{Introduction}\label{sec:introduction}

The idea that our universe might be a 4-dimensional hypersurface
embedded in a higher-dimensional manifold is an old one with a
long history, as well as the subject of a considerable amount of
contemporary interest.  The primordial impetus for this line of
study comes from the work of Kaluza \cite{Kal21}, who showed that
one can obtain a classical unification of gravity and
electromagnetism by adding an extra dimension to spacetime (1921);
and Klein \cite{Kle26}, who suggested that extra dimensions have a
circular topology of very small radius and are hence unobservable
(1926).  The latter idea, the so-called ``compactification''
paradigm, came to dominate most approaches to higher-dimensional
physics, the most notable of which was early superstring theory.
However, a number of papers have appeared in the intervening years
that do not assume extra dimensions with compact topologies; early
examples include the works of Joseph \cite{Jos62}, Akama
\cite{Aka82}, Rubakov \& Shaposhnikov \cite{Rub83}, Visser
\cite{Vis85}, Gibbons \& Wiltshire \cite{Gib87}, and Antoniadis
\cite{Ant00}. A systematic and independent approach to the
5-dimensional, non-compact Kaluza-Klein scenario, known as
Space-Time-Matter (STM) theory, followed
\cite{Wes92,Wes96,Ove97,Wes99}. Then, in 1996 Horava \& Witten
showed that the compactification paradigm was not a prerequisite
of string theory with their discovery of an 11-dimensional theory
on the orbifold $\mathbb{R}^{10} \times S^1 / \mathbb{Z}_2$, which
is related to the 10-dimensional $E_8 \times E_8$ heterotic string
via dualities \cite{Hor96}. In this theory, standard model
interactions are confined to a 3-brane, on which the endpoints of
open strings reside, while gravitation propagates in the
higher-dimensional bulk. This situation has come to be known as
the ``braneworld scenario.''  The works of Arkani-Hamed \emph{et
al.} \cite{Ark98a,Ant98,Ark98b} and Randall \& Sundrum (RS)
\cite{Ran99a,Ran99b}, which used non-compact extra dimensions to
address the hierarchy problem of particle physics and demonstrated
that the graviton ground state can be localized on a 3-brane in 5
dimensions, won a large following for the braneworld scenario.  A
virtual flood of papers dealing with non-compact,
higher-dimensional models of the universe soon followed, including
works dealing with ``thick'' branes --- that is, 3-branes with
finite extra-dimensional width \cite{DeW99,Csa00}.

On the other hand, the abstract problem of embedding an
$n$-dimensional (pseudo-) Riemannian manifold $\Sigma_0$ in a
higher-dimensional space $M$ also has a rich pedigree\footnote{An
extensive bibliography is available from ref.~\cite{Pav00}.}. Soon
after Riemann published the theory of intrinsically-defined
abstract manifolds in 1868 \cite{Rie68}, Schl\"afli considered the
problem of how to locally embed such manifolds in Euclidean space
\cite{Sch71}. He conjectured that the maximum number of extra
dimensions necessary for a local embedding was $\tfrac{1}{2}
n(n-1)$.  In 1926, Janet provided a partial proof of the
conjecture for $n =2$ using power series methods \cite{Jan26}.
That result was soon generalized to arbitrary $n$ by Cartan
\cite{Car27}. The proof was completed by Burstin \cite{Bur31}, who
demonstrated that the Gauss-Codazzi-Ricci equations were the
integrability conditions of the embedding.  A related embedding
problem was first considered by Campbell in 1926 \cite{Cam26}: How
many extra dimensions are required to locally embed $\Sigma_0$ in
a Ricci-flat space; i.e., a higher-dimensional vacuum spacetime?
He proposed that the answer was one, which was later proved by
Magaard \cite{Mag63}. The Campbell-Magaard theorem has recently
been extended to include cases where the higher-dimensional space
has a nonzero cosmological constant \cite{And01a,Dah01a}, is
sourced by a scalar field \cite{And01b}, and has an arbitrary
non-degenerate Ricci tensor \cite{Dah01b}.  It is important to
note that the results discussed thus far are local; the problem of
global embedding arbitrary sub-manifolds is more difficult.  In
1956, Nash showed that the minimum number of extra dimension
required to embed $\Sigma_0$ in Euclidean space is $3n(n+3)/2$ if
$\Sigma_0$ is compact --- it increases to $3n(n+1)(n+3)/2$ if
$\Sigma_0$ is non-compact \cite{Nas56}.  Global results for
metrics with indefinite signature were later obtained by Clarke
\cite{Cla70} and Greene \cite{Gre70}.

It is obvious that there is a significant interplay between
higher-dimensional theories of the universe and mathematical
embedding theorems.  Specifically, the Campbell-Magaard theorem
and its variants would seem to be of particular importance to
higher-dimensional physics because the embedding space in such
theories is usually specified solely by its stress-energy
--- and hence Ricci --- tensor.  Yet this theorem
has received only moderate coverage in the literature \cite{Rom96,Mai95,%
Lid97a,Lid97b}.  One of the primary motivations of this paper is
to give an introduction to, and a non-rigourous proof of, the
Campbell-Magaard theorem (Sec.~\ref{sec:general}).  In the
process, we will see how the theorem can be altered to fit any
number of circumstances.  For example, we will see that it is
possible to successfully embed $\Sigma_0$ in $M$ with arbitrary
extrinsic (as opposed to intrinsic) curvature.  We will then
discuss general geometric properties of several 5-dimensional
theories --- STM and the thin/thick braneworld scenarios --- as
well as the application of the Campbell-Magaard theorem to each
(Sec.~\ref{sec:applications}).  Our discussion will demonstrate
that these seemingly disparate models actually have a lot in
common.  We will also generalize the $(4+1)$-splitting of the
5-dimensional geodesic equation found in ref.~\cite{Sea02} to
analyze the motion of test observers in each scenario.  We will
see that in all the cases considered, the physical demand that
observer trajectories are confined to $\Sigma_0$ makes it
impossible to successfully implement a 5-dimensional embedding of
arbitrary 4-manifolds.  However, the restrictions placed on the
geometry of embedded spacetimes by this requirement are different
for each scenario. Finally, we will illustrate several aspects of
the proceeding discussion by considering the $\mathbb{Z}_2$
symmetric embedding of standard radiation-dominated cosmologies
in a 5-dimensional vacuum manifold. This embedding is obtained
from a known solution of STM theory found in ref.~\cite{Liu01}.
We will see that observers can be confined to $\Sigma_0$ in this
case, but that their equilibrium is unstable.
Sec.~\ref{sec:conclusions} is reserved for a summary of our
results.

\paragraph*{Conventions.}
Uppercase Latin indices run from $0 \ldots n$, lowercase Greek
indices run $0 \ldots (n-1)$, and lowercase Latin indices run $1
\ldots \frac{3}{2} n(n+1)$.  Square brackets on indices indicate
anti-symmetrization.  The $(n+1)$-dimensional covariant
derivative will be indicated by $\nabla_A$, while the
$n$-dimensional covariant derivative will be indicated with a
semicolon. Curvature tensors in $(n+1)$ dimensions are
distinguished from their $n$-dimensional counterparts with a hat;
i.e.~$\hat{R}_{AB}$ versus $R_{\alpha\beta}$.  We choose units
where $c = 1$.  The coupling constant between the $n$-dimensional
Einstein and stress-energy tensors is $\kappa_n^2$ such that
$G_{\alpha\beta} = \kappa_n^2 T_{\alpha\beta}$.

\section{The general embedding
problem}\label{sec:general}

In this section, we discuss the general problem of embedding an
$n$-dimensional spacetime in an $(n+1)$-dimensional manifold.
Certain aspects of our discussion may be familiar to some readers
from other contexts.  Our philosophy will to be to ignore such
previous knowledge and start from basic geometric principles.  In
this way, we hope to emphasize various points that play an
important role in the applications discussed in
Sec.~\ref{sec:applications}.

\subsection{Geometric Construction}\label{sec:geometry}

We will be primarily concerned with an $(n+1)$-dimensional
manifold $(M,g_{AB})$ on which we place a coordinate system $x
\equiv \{ x^A \}$.  In our working, we will allow for two
possibilities: either there is one timelike and $n$ spacelike
directions tangent to $M$, \emph {or} there are two timelike and
$(n-1)$ spacelike directions tangent to $M$.  We introduce a
scalar function
\begin{equation}
    \ell = \ell(x),
\end{equation}
which defines our foliation of the higher-dimensional manifold
with the hypersurfaces given by $\ell$ = constant, denoted by
$\Sigma_\ell$.  If there is only one timelike direction tangent
to $M$, we assume that the vector field $n^A$ normal to
$\Sigma_\ell$ is spacelike.  If there are two timelike
directions, we take the unit normal to be timelike.  In either
case, the space tangent to a given $\Sigma_\ell$ hypersurface
contains one timelike and $(n-1)$ spacelike directions.  That is,
each $\Sigma_\ell$ hypersurface corresponds to an $n$-dimensional
Lorentzian spacetime.  The normal vector to the $\Sigma_\ell$
slicing is given by
\begin{equation}
    n_A = \varepsilon \Phi \, \di_A \ell, \quad n^A n_A = \varepsilon.
\end{equation}
Here, $\varepsilon = \pm 1$.  The scalar $\Phi$ which normalizes
$n^A$ is known as the lapse function.  We define the projection
tensor as
\begin{equation}
\label{induced metric}
    h_{AB} = g_{AB} - \varepsilon n_A n_B.
\end{equation}
This tensor is symmetric ($h_{AB} = h_{BA}$) and orthogonal to
$n_A$.  We place an $n$-dimensional coordinate system on each of
the $\Sigma_\ell$ hypersurfaces $y \equiv \{ y^\alpha \}$.  The
$n$ basis vectors
\begin{equation}
    e^A_\alpha = \frac{\di x^A}{\di y^\alpha}, \quad n_A
    e^A_\alpha = 0
\end{equation}
are by definition tangent to the $\Sigma_\ell$ hypersurfaces and
orthogonal to $n^A$.  It is easy to see that $e^A_\alpha$ behaves
as a vector under coordinate transformations on $M$ [$\phi:x
\rightarrow \bar{x} (x)$] and a one-form under coordinate
transformations on $\Sigma_\ell$ [$\psi:y \rightarrow
\bar{y}(y)$]. We can use these basis vectors to project
higher-dimensional objects onto $\Sigma_\ell$ hypersurfaces.  For
example, for an arbitrary one-form on $M$ we have
\begin{equation}
    T_\alpha = e_\alpha^A T_A.
\end{equation}
Here $T_\alpha$ is said to be the projection of $T_A$ onto
$\Sigma_\ell$.  Clearly $T_\alpha$ behaves as a scalar under
$\phi$ and a one-form under $\psi$.  The induced metric on the
$\Sigma_\ell$ hypersurfaces is given by
\begin{equation}
    h_{\alpha\beta} = e^A_\alpha e^B_\beta g_{AB} = e^A_\alpha
    e^B_\beta h_{AB}.
\end{equation}
Just like $g_{AB}$, the induced metric has an inverse:
\begin{equation}
    h^{\alpha\gamma} h_{\gamma\beta} = \delta^{\alpha}_{\beta}.
\end{equation}
The induced metric and its inverse can be used to raise and lower
the indices of tensors tangent to $\Sigma_\ell$, and change the
position of the spacetime index of the $e^A_\alpha$ basis
vectors.  This implies
\begin{equation}
    e_A^\alpha e^A_\beta = \delta^\alpha_\beta.
\end{equation}
Also note that since $h_{AB}$ is entirely orthogonal to $n^A$, we
can express it as
\begin{equation}\label{induced decomposition}
    h_{AB} = h_{\alpha\beta} e^\alpha_A e^\beta_B.
\end{equation}
At this juncture, it is convenient to introduce our definition of
the extrinsic curvature $K_{\alpha\beta}$ of the $\Sigma_\ell$
hypersurfaces:
\begin{equation}\label{extrinsic def}
    K_{\alpha\beta} = e^A_\alpha e^B_\beta \nabla_A n_B =
    \tfrac{1}{2} e^A_\alpha e^B_\beta \pounds_n h_{AB}.
\end{equation}
Note that the extrinsic curvature is symmetric ($K_{\alpha\beta}
= K_{\beta\alpha}$).  It may be thought of as the derivative of
the induced metric in the normal direction.  This $n$--tensor
will appear often in what follows.

Finally, we note that $\{ y, \ell \}$ defines an alternative
coordinate system to $x$ on $M$.  The appropriate diffeomorphism
is
\begin{equation}\label{diffeo}
    dx^A = e_\alpha^A dy^\alpha + \ell^A d\ell,
\end{equation}
where
\begin{equation}\label{flow of l vector}
    \ell^A =  \left( \frac{\di x^A}{\di \ell} \right)_{y^\alpha =
    \mathrm{const.}}
\end{equation}
is the vector tangent to lines of constant $y^\alpha$.  We can
always decompose 5D vectors into the sum of a part tangent to
$\Sigma_\ell$ and a part normal to $\Sigma_\ell$.  For $\ell^A$
we write
\begin{equation}\label{l vector}
    \ell^A = N^\alpha e_\alpha^A + \Phi n^A.
\end{equation}
This is consistent with $\ell^A \di_A \ell = 1$, which is
required by the definition of $\ell^A$, and the definition of
$n^A$.  The $n$--vector $N^\alpha$ is the shift vector, which
describes how the $y^\alpha$ coordinate system changes as one
moves from a given $\Sigma_\ell$ hypersurface to another. Using
our formulae for $dx^A$ and $\ell^A$, we can write the 5D line
element as
\begin{eqnarray}\nonumber
    ds_{(5)}^2 & = & g_{AB} \, dx^A dx^B \\ & = & h_{\alpha\beta}
    (dy^\alpha + N^\alpha d\ell) (dy^\beta + N^\beta d\ell)
    \nonumber \\ & & + \varepsilon \Phi^2 d\ell^2.
\end{eqnarray}
This reduces to $d\mathcal{S}^2 = h_{\alpha\beta} dy^\alpha
dy^\beta$ if $d\ell = 0$, a case of considerable physical
interest.

\subsection{Decomposition of the higher-dimensional field
equations}\label{sec:decomposition}

In this section, we describe how $n$-dimensional field equations
on each of the $\Sigma_\ell$ hypersurfaces can be derived, given
that the $(n+1)$-dimensional field equations are
\begin{equation}\label{5d field eqns}
    \hat{R}_{AB} = \lambda g_{AB}, \quad \lambda \equiv
    \frac{2\Lambda}{1-n}.
\end{equation}
Here $\Lambda$ is the ``bulk'' cosmological constant, which may
be set to zero if desired.  In what follows, we will extent the
4-dimensional usage and call manifolds satisfying equation
(\ref{5d field eqns}) Einstein spaces.

Our starting point is the Gauss-Codazzi equations.  On each of
the $\Sigma_\ell$ hypersurfaces these read
\begin{subequations}\label{gauss codazzi}
\begin{eqnarray}
    \hat{R}_{ABCD} e^A_\alpha e^B_\beta e^C_\gamma e^D_\delta & =
    & R_{\alpha\beta\gamma\delta} +2 \varepsilon
    K_{\alpha [ \delta} K_{\gamma ] \beta}, \\
    \hat{R}_{MABC} n^M  e^A_\alpha e^B_\beta e^C_\gamma & = &
    2 K_{\alpha [\beta ; \gamma]}.
\end{eqnarray}
\end{subequations}
These need to be combined with the following expression for the
higher-dimensional Ricci tensor:
\begin{equation}\label{ricci decomposition}
    \hat{R}_{AB} = ( h^{\mu\nu} e_\mu^M e_\nu^N + \varepsilon
    n^M n^N ) \hat{R}_{AMBN}.
\end{equation}
The $\tfrac{1}{2} (n+1) (n+2)$ separate equations for the
components of $\hat{R}_{AB}$ may be broken up into three sets by
considering the following contractions of equation (\ref{5d field
eqns}):
\begin{equation}\label{projections}
\begin{array}{rclcr}
    \hat{R}_{AB} e^A_\alpha e^B_\beta & = & \lambda h_{\alpha\beta},
    & & \text{$\tfrac{1}{2}n(n+1)$ equations}, \\
    \hat{R}_{AB} e^A_\alpha n^B & = & 0,
    & & \text{$n$ equations}, \\
    \hat{R}_{AB} n^A n^B & = & \varepsilon \lambda, &
    \, & + \text{1 equation,} \\
    & & & & \overline{\text{$\tfrac{1}{2} (n+1) (n+2)$ equations.}}
\end{array}
\end{equation}
Putting (\ref{ricci decomposition}) into (\ref{projections}) and
making use of (\ref{gauss codazzi}) yields the following
formulae:
\begin{subequations}\label{4d field eqns}
\begin{eqnarray}\label{4d field eqns 1}
    R_{\alpha\beta} & = & \lambda h_{\alpha\beta} - \varepsilon
    [ E_{\alpha\beta} + K_{\alpha}{}^{\mu} ( K_{\beta\mu} -
    K h_{\beta\mu} ) ], \\ \label{4d field eqns 2} 0 & = &
    (K^{\alpha\beta} - h^{\alpha\beta}
    K)_{;\alpha}, \\ \label{4d field eqns 3} \varepsilon \lambda
    & = & E_{\mu\nu} h^{\mu\nu}.
\end{eqnarray}
\end{subequations}
In writing down these results, we have made the following
definitions:
\begin{equation}
    K \equiv h^{\alpha\beta} K_{\alpha\beta},
\end{equation}
\begin{equation}\label{weyl def}
    E_{\alpha\beta} \equiv \hat{R}_{MANB} n^M e^A_\alpha n^N
    e^B_\beta, \quad E_{\alpha\beta} = E_{\beta\alpha}.
\end{equation}
The Einstein tensor $G_{\alpha\beta} = R_{\alpha\beta} -
\tfrac{1}{2} g_{\alpha\beta} R$ on a given $\Sigma_\ell$
hypersurface is given by
\begin{eqnarray}\nonumber
    G^{\alpha\beta} & = & - \varepsilon \left( E^{\alpha\beta} +
    K^{\alpha}{}_{\mu} P^{\mu\beta} - \tfrac{1}{2}
    h^{\alpha\beta} K^{\mu\nu} P_{\mu\nu} \right) \\ \label{4d einstein}
    & & - \tfrac{1}{2} \lambda (n-3) h^{\alpha\beta},
\end{eqnarray}
where we have defined the (conserved) tensor
\begin{equation}
    P_{\alpha\beta} \equiv K_{\alpha\beta} - h_{\alpha\beta} K,
    \quad P^{\alpha\beta}{}_{;\beta} = 0.
\end{equation}
Some will recognize this as algebraically equivalent the momentum
conjugate to the induced metric in the ADM Hamiltonian
formulation of general relativity.\footnote{But we should keep in
mind that the direction orthogonal to $\Sigma_\ell$ is not
necessarily timelike, so $P_{\alpha\beta}$ cannot formally be
identified with a canonical momentum variable in the Hamiltonian
sense.} The condition that the Einstein 4-tensor has zero
divergence translates into a condition satisfied by
$E_{\alpha\beta}$:
\begin{equation}\label{weyl field eqn}
    E^{\alpha\beta}{}_{;\beta} = \varepsilon ( K_{\mu\nu} K^{\mu
    \nu ; \alpha} - K^{\mu\beta} K^{\alpha}{}_{\mu ; \beta}).
\end{equation}
In deriving this formula, we have made use of (\ref{4d field eqns
2}).  Equations (\ref{4d field eqns}) are the field equations
governing $n$-dimensional curvature quantities on the
$\Sigma_\ell$ hypersurfaces.  We will study the properties of
these field equations in the next section.

Before moving on, we note that it is possible to solve equation
(\ref{4d field eqns 1}) for $E_{\alpha\beta}$ and substitute the
result into (\ref{4d field eqns 3}) to get
\begin{equation}\label{adm constraint}
    (n-1)\lambda = R + \varepsilon( K^{\mu\nu} K_{\mu\nu} - K^2
    ).
\end{equation}
Taken together, equations (\ref{4d field eqns 2}) and (\ref{adm
constraint}) are $(n+1)$-dimensional generalizations of the
well-known Hamiltonian constraints, familiar to those who work
with numerical relativity or initial-value problems in $(3+1)$
dimensions.  So these formulae have been written down before, and
similar equations have been used to analyze the RS scenario by
Shiromizu \emph{et al.}~\cite{Shi99}, though not in the context of
the theorem we are about to describe.

\subsection{The generalized Campbell-Magaard
theorem}\label{sec:campbell}

In this section, we discuss generic features of the embedding
problem and outline a proof of the Campbell-Magaard theorem based
on the formulae derived above. Technical comments can be found in
refs.~\cite{Mag63,And01a,Dah01a,And01b,Dah01b}, but here we wish
to provide a physically-motivated argument.

In the previous section, we derived field equations for the
$n$-dimensional tensors --- which can be thought of as three
spin-2 fields --- defined on each of the $\Sigma_\ell$
hypersurfaces:
\begin{equation}\label{4-tensors}
    h_{\alpha\beta} (y,\ell), \quad K_{\alpha\beta} (y,\ell),
    \quad E_{\alpha\beta} (y,\ell).
\end{equation}
Each of the tensors is symmetric, so there are $3 \times
\frac{1}{2} n(n+1)$ independent dynamical quantities governed by
the field equations (\ref{4d field eqns}).  For book-keeping
purposes, we can organize these into an $n_\text{dyn} =
\tfrac{3}{2} n(n+1)$-dimensional super-vector $\Psi^a =
\Psi^a(y,\ell)$. Now, the field equations (\ref{4d field eqns})
contain no derivatives of the tensors (\ref{4-tensors}) with
respect to $\ell$.  This means that the components
$\Psi^a(y,\ell)$ must satisfy (\ref{4d field eqns}) and
(\ref{weyl field eqn}) for \emph{each and every} value of $\ell$.
In an alternative language, the field equations on $\Sigma_\ell$
are ``conserved'' as we move from hypersurface to hypersurface.
That is, the field equations (\ref{4d field eqns}) in
$(n+1)$-dimensions are, in the Hamiltonian sense,
\emph{constraint equations}.  While this is important from the
formal viewpoint, in means that equations (\ref{5d field eqns})
tell us nothing about how $\Psi^a$ varies with $\ell$.  Equations
governing the $\ell$--evolution of $\Psi^a$ may be derived in a
number of equivalent ways.  These include isolating
$\ell$-derivatives in the expansion of the Bianchi identities
$\nabla_A G^{AB} = 0$; direct construction of the Lie derivatives
of $h_{AB}$ and $K_{AB} = h_A{}^C \nabla_C n_B$ with respect to
$\ell^A$; and formally re-expressing the gravitational Lagrangian
as a Hamiltonian (with $\ell$ playing the role of time) to obtain
the equations of motion.  Because the derivation of $\di_\ell
\Psi^a$ is tedious and not really germane to our discussion, we
will omit it from our considerations and turn to the more
important problem of embedding.

Essentially, our goal is to find a solution of the
higher-dimensional field equations (\ref{5d field eqns}) such
that \emph{one} hypersurface $\Sigma_0$ in the $\Sigma_\ell$
foliation has ``desirable'' geometrical properties.  For example,
we may want to completely specify the induced metric on, and
hence the intrinsic geometry of, $\Sigma_0$.  Without loss of
generality, we can assume that the hypersurface of interest is at
$\ell = 0$. Then to successfully embed $\Sigma_0$ in $M$, we need
to do two things:
\begin{enumerate}
    \item Solve the constraint equations (\ref{4d field eqns}) on
    $\Sigma_0$ for $\Psi^a(y,0)$ such that $\Sigma_0$ has the desired
    properties (this involves physics).
    \item Obtain the solution for $\Psi^a(y,\ell)$ in the bulk
    (i.e. for $\ell \ne 0$) using the evolution equations $\di_\ell
    \Psi^a$ (this is mainly mathematics).
\end{enumerate}
To prove the Campbell-Magaard theorem one has to show that Step 1
is possible for arbitrary choices of $h_{\alpha\beta}$ on
$\Sigma_0$, and one also needs to show that the bulk solution for
$\Psi^a$ obtained in Step 2 preserves the equations of constraint
on $\Sigma_\ell \ne \Sigma_0$.  The latter requirement is
necessary because if the constraints are not conserved, the
higher-dimensional field equations will not hold away from
$\Sigma_0$. This issue has been considered by several authors,
who have derived evolution equations for $\Psi^a$ and
demonstrated that the constraints are conserved in quite general
$(n+1)$-dimensional manifolds \cite{And01a,Dah01a,Dah01b}. Rather
than dwell on this well-understood point, we will concentrate on
the $n$-dimensional field equations on $\Sigma_0$ and assume
that, given $\Psi^a(y,0)$, then the rest of $(n+1)$-dimensional
geometry can be generated using evolution equations, with the
resulting higher-dimensional metric satisfying the appropriate
field equations.  However, we expect that the practical
implementation of the formal embedding procedure given above will
be fraught with the same type of computational difficulties
associated with the initial-value problem in ordinary general
relativity, and is hence a nontrivial exercise.

Now, there are $n_\text{cons} = \tfrac{1}{2} (n+1) (n+2)$
constraint equations on $\Sigma_0$. For $n \ge 2$ we see that
$\dim \Psi^a = n_\text{dyn}$ is greater than $n_\text{cons}$,
which means that our system is underdetermined. Therefore, we may
freely specify the functional dependence of $n_\text{free} = n^2
-1$ components of $\Psi^a(y,0)$.  This freedom is at the heart of
the Campbell-Magaard theorem.  Since $n_\text{free}$ is greater
than the number of independent components of $h_{\alpha\beta}$
for $n \ge 2$, we can choose the line element on $\Sigma_0$ to
correspond to any $n$-dimensional Lorentzian manifold and still
satisfy the constraint equations.  This completes the proof of
the theorem, \emph{any $n$-dimensional manifold can be locally
embedded in an $(n+1)$-dimensional Einstein space}.

We make a few comments before moving on: First, it is equally
valid to fix $K_{\alpha\beta}(y,0)$
--- or even $E_{\alpha\beta}(y,0)$
--- instead of $h_{\alpha\beta}(y,0)$. That is, instead of
specifying the \emph{intrinsic} curvature of $\Sigma_0$, one
could arbitrarily choose the \emph{extrinsic} curvature. However,
it is obvious that we cannot arbitrarily specify \emph{both} the
induced metric and extrinsic curvature of $\Sigma_0$, \emph{and}
still solve the constraints: there are simply not enough degrees
of freedom. We will see in Sec.~\ref{sec:applications} that there
are several scenarios where we will want to set
$K_{\alpha\beta}(y,0) = 0$, but that in doing so we will severely
restrict the intrinsic geometry of $\Sigma_0$.

Second, one might legitimately wonder about the
$n_\mathrm{residual} = n_\mathrm{dyn} - n_\mathrm{cons} -
\tfrac{1}{2} n (n+1) = \tfrac{1}{2}(n+1)(n-2)$ degrees of freedom
in $\Psi^a(y,0)$ ``left over'' after the constraints are imposed
and the induced metric is selected.  What role do these play in
the embedding?  The existence of some degree of arbitrariness in
$\Psi^a(y,0)$, which essentially comprises the initial data for a
Cauchy problem, would seem to suggest that when we choose an
induced metric on $\Sigma_0$ we do not uniquely fix the
properties of the bulk.  In other words, we can apparently embed
the same $n$-manifold in different Einstein spaces.  For example,
in Sec.~\ref{sec:example} we will see how a 4-dimensional
radiation-dominated cosmological model can be embedded in a
5-dimensional bulk with $\hat{R}_{AB} = 0$ and $\hat{R}_{ABCD}
\ne 0$. However, in refs.~\cite{Pon88,Sea02b} it was demonstrated
that the same 4-manifold can been embedded in 5-dimensional
Minkowski space. These results confirm that, in general, the
structure of $M$ is not determined uniquely by the intrinsic
geometry of $\Sigma_0$.

Third, we reiterate that the Campbell-Magaard theorem is a local
result.  It does not state that it is possible to embed
$\Sigma_0$ with arbitrary global topology into an
$(n+1)$-dimensional Einstein space.  As far as we know, the issue
of how many extra dimensions are required for a global embedding
of $\Sigma_0$ an Einstein space is an open question.

\section{Applications}\label{sec:applications}

We now want to apply the formalism discussed in
Sec.~\ref{sec:general} to three higher-dimensional theories of
our universe: STM theory, the RS thin braneworld scenario, and
the thick braneworld scenario. The common feature of these models
is that our $(3+1)$-dimensional spacetime is viewed as a
hypersurface in a 5-dimensional manifold.  For obvious reasons,
we take $n = 4$ when using equations from Sec.~\ref{sec:general}
in the current discussion. Then, each $\Sigma_\ell$ hypersurface
can be identified with a 4-dimensional embedded spacetime.

\subsection{Space-Time-Matter theory}

STM theory predates the current flurry of interest in non-compact
5-dimensional models spurred by the work of RS. Essentially, it
is a \emph{minimal} model that assumes the 5-dimensional manifold
to be devoid of matter.  That is, the 5-manifold is Ricci-flat.
The field equations of this theory are then given by equations
(\ref{4d field eqns}) with $\lambda = 0$. Despite the fact that
we have $\hat{G}_{AB} = 0$, the Einstein tensor on each of the
embedded spacetimes is non-trivial and is given by equation
(\ref{4d einstein}) with vanishing bulk cosmological constant:
\begin{equation}\label{stm einstein}
    G^{\alpha\beta} = - \varepsilon \left( E^{\alpha\beta} +
    K^{\alpha}{}_{\mu} P^{\mu\beta} - \tfrac{1}{2}
    h^{\alpha\beta} K^{\mu\nu} P_{\mu\nu} \right).
\end{equation}
Matter enters into STM theory when we consider an observer who is
capable of performing experiments that measure the 4-metric
$h_{\alpha\beta}$ or Einstein tensor $G_{\alpha\beta}$ in some
neighbourhood of their position, yet is ignorant of the dimension
transverse to his spacetime, the 5-metric $g_{AB}$ and
5-dimensional curvature tensors. For general situations, such an
observer will discover that his universe is curved, and that the
local Einstein tensor is given by (\ref{stm einstein}). Now, if
this person believes in the Einstein equations $G_{\alpha\beta} =
\kappa_4^2 T_{\alpha\beta}$, he will be forced to conclude that
the spacetime around him is filled with some type of matter
field. This is a somewhat radical departure from the usual point
of view that the stress-energy distribution of matter fields acts
as the source of the curvature of the universe.  In the STM
picture, the shape of the $\Sigma_\ell$ hypersurfaces plus the
5-dimensional Ricci-flat geometry fixes the matter distribution.
It is for this reason that STM theory is sometimes called
induced-matter theory: the matter content of the universe is
induced from higher-dimensional geometry.  This geometrization of
matter is the primary motivation for studying STM theory.  (For
an in-depth review of this formalism, see ref.~\cite{Wes99}.)

When applied to STM theory, the Campbell-Magaard theorem says
that it is possible to specify the form of $h_{\alpha\beta}$ on
one of the embedded spacetimes, denoted by $\Sigma_0$. In other
words, \emph{we can take any known $(3+1)$-dimensional solution
$h_{\alpha\beta}$ of the Einstein equations for matter with
stress-energy tensor $T_{\alpha\beta}$ and embed it on a
hypersurface in the STM scenario}.  The stress-energy tensor of
the induced matter on that hypersurface $\Sigma_0$ will
necessarily match that of the $(3+1)$-dimensional solution.
However, there is no guarantee that the induced matter on any of
the other spacetimes will have the same properties.

We now wish to expand the discussion to include the issue of
observer trajectories in STM theory.  To do this, we will need
the covariant decomposition of the equation of motion for test
observers into parts describing tangential and orthogonal
accelerations; this is accomplished in Appendix \ref{sec:EOM}.
The salient result from that discussion is equation (\ref{extra
EOM}), which governs motion in the $\ell$ direction.  We now
analyze this formula for three different cases that may apply to
a given $\Sigma_\ell$ 4-surface in an STM manifold.  (The various
quantities appearing in the following are defined in Appendix
\ref{sec:EOM}.)

\begin{enumerate}

\item \emph{$K_{\alpha\beta} \ne 0$ and ${\mathfrak{F}} = 0$}. A sub-class of
this case has $f^\beta = 0$, which corresponds to freely-falling
observers.  We cannot have $\ell =$ constant as a solution of the
$\ell$ equation of motion (\ref{extra EOM}) in this case, so
observers cannot live on a single hypersurface.  Therefore, if we
construct a Ricci-flat 5-dimensional manifold in which a
particular solution of general relativity is embedded on
$\Sigma_0$, and we put an observer on that hypersurface, then
that observer will inevitability move in the $\ell$ direction.
Therefore, the properties of the induced matter that the observer
measures may match the predictions of general relativity for a
brief period of time, but this will not be true in the long run.
Therefore, STM theory predicts observable departures from general
relativity.

\item \emph{$K_{\alpha\beta} = 0$ and ${\mathfrak{F}} = 0$}.  Again, this case
includes freely-falling observers.  Here, we can solve equation
(\ref{extra EOM}) with $d\ell / d\lambda = 0$. That is, if a
particular hypersurface $\Sigma_0$ has vanishing extrinsic
curvature, then we can have observers with trajectories contained
entirely within that spacetime, provided that ${\mathfrak{F}} =
0$. Such hypersurfaces are called geodesically complete because
every geodesic on $\Sigma_0$ is also a geodesic of $M$
\cite{Sea02,Ish01}. If we put $K_{\alpha\beta} = 0$ into equation
(\ref{stm einstein}), then we get the Einstein tensor on
$\Sigma_\ell$ as
\begin{equation}
    G_{\alpha\beta} = \kappa_4^2 T_{\alpha\beta} = -
    \varepsilon E_{\alpha\beta} \,\, \Rightarrow \,\,
    T^\alpha{}_\alpha = 0.
\end{equation}
This says that the stress-energy tensor of the induced matter
associated with geodesically complete spacetimes must have zero
trace. Assuming that the stress-energy tensor may expressed as
that of a perfect fluid, this implies a radiation-like equation
of state. Hence, it is impossible to embed an \emph{arbitrary}
spacetime in a 5-dimensional vacuum such that it is geodesically
complete. This is not surprising, since we have already seen in
Sec.~\ref{sec:campbell} that we cannot use the Campbell-Magaard
theorem to choose both $h_{\alpha\beta}$ and $K_{\alpha\beta}$ on
$\Sigma_0$ --- we have the freedom to specify one or the other,
but not both.  If we do demand that test observers are
gravitationally confined to $\Sigma_0$, we place strong
restrictions on the geometry of the embedded spacetime.

\item \emph{$K_{\alpha\beta} \ne 0$ and ${\mathfrak{F}} = -
K_{\alpha\beta} u^\alpha u^\beta$}.  In this case, we can solve
(\ref{extra EOM}) with $d\ell/d\lambda = 0$ and hence have
observers confined to the $\Sigma_0$ spacetime.  It was shown in
ref.~\cite{Sea02} that ${\mathfrak{F}} = - K_{\alpha\beta}
u^\alpha u^\beta$ is merely the higher-dimensional generalization
of the centripetal acceleration familiar from elementary
mechanics. Since we do not demand $K_{\alpha\beta} = 0$ in this
case, we can apply the Campbell-Magaard theorem and have any type
of induced matter on $\Sigma_0$.  However, the price to be paid
for this is the inclusion of a non-gravitational centripetal
confining force, whose origin is obscure.

\end{enumerate}

To summarize, we have shown that the Campbell-Magaard theorem
guarantees that we can embed any solution of general relativity
on a spacetime hypersurface $\Sigma_0$ within the 5D manifold
postulated by STM theory.  However, in general situations
observer trajectories will not be confined to $\Sigma_0$. The
exception to this is when $\Sigma_0$ has $K_{\alpha\beta} = 0$:
then observers with ${\mathfrak{F}} = 0$ remain on $\Sigma_0$
given the initial condition $\xi(\lambda_0) = 0$.  However,
$K_{\alpha\beta} = 0$ places a restriction on the induced matter,
namely $T^\alpha{}_\alpha = 0$.  Finally, if observers are
subject to a non-gravitational force such that ${\mathfrak{F}} =
-K_{\alpha\beta} u^\alpha u^\beta$, then they can be confined to
general $\Sigma_0$.  The source of this centripetal confining
force is not clear.

\subsection{The thin braneworld scenario}

We now move on to the widely-referenced thin braneworld scenarios
proposed by Randall \& Sundrum (RS).  There are actually two
different versions of the thin braneworld picture: the so-called
RSI and RSII models. In both situations, one begins by assuming a
5-dimensional manifold with a non-zero cosmological constant
$\Lambda$, which is often taken to be negative; i.e., the bulk is
AdS${}_5$. The $\Sigma_0$ hypersurface located at $\ell = 0$
represents a domain wall across which the normal derivative of
the metric (the extrinsic curvature) is discontinuous. Those
familiar with the thin-shell formalism in general relativity will
realize that such a discontinuity implies that there is a thin
4-dimensional matter configuration living on $\Sigma_0$.  The
motivation for such a geometrical setup comes from the Horava \&
Witten theory mentioned in Sec.~\ref{sec:introduction}, where
standard model fields are effectively confined to a 3-brane in a
higher-dimensional manifold. In the RS models, the distributional
matter configuration corresponds to these matter fields.  Now, if
we stop here we have described the salient features of the RSII
model.  This scenario has drawn considerable interest in the
literature, because for certain solutions the Kaluza-Klein
spectrum of the graviton is such that Newton's $1/r^2$ law of
gravitation is unchanged over astronomical length scales.  By
contrast, the RSI model differs from RSII by the addition of a
second 3-brane located at some $\ell \ne 0$. The motivation for
the addition of the second brane comes from a possible solution
of the hierarchy problem, which involves the disparity in size
between the characteristic energies of quantum gravity and
electroweak interactions.  The idea is that the characteristic
lengths, and hence energy scales, on the 3-branes are
exponentially related by the intervening AdS${}_5$ space.  In
what follows, we will concentrate mostly on the RSII scenario,
although many of our comments can be applied to RSI.

When we apply the standard Israel junction conditions
\cite{Isr66} to RSII, we find that the induced metric on the
$\Sigma_\ell$ hypersurfaces must be continuous:
\begin{equation}
    [h_{\alpha\beta}] = 0.
\end{equation}
We adopt the common notation that $X^\pm \equiv \lim_{\ell
\rightarrow 0^\pm} X$ and $[X] = X^+ - X^-$.  In addition, the
Einstein tensor of the bulk is given by
\begin{equation}
    \hat{G}_{AB} = \Lambda g_{AB} + \kappa_5^2 T^{(\Sigma)}_{AB}, \quad
    T^{(\Sigma)}_{AB} = \delta ( \ell ) \, S_{\alpha
    \beta} e^\alpha_A e^\beta_B.
\end{equation}
Here, the 4-tensor $S_{\alpha\beta}$ is defined as
\begin{equation}
    [K_{\alpha\beta}] \equiv  -\kappa_5^2 \varepsilon
    \left( S_{\alpha\beta} - \tfrac{1}{3} S h_{\alpha\beta}
    \right),\label{K-S relation}
\end{equation}
where $S = h^{\mu\nu} S_{\mu\nu}$.  The interpretation is that
$S_{\alpha\beta}$ is the stress-energy tensor of the standard
models fields on the brane. To proceed further, we need to invoke
another assumption of the RS model, namely the $\mathbb{Z}_2$
symmetry.  This \emph{ansatz} essentially states that the
geometry on one side of the brane is the mirror image of the
geometry on the other side. In practical terms, it implies
\begin{equation}
    K^+_{\alpha\beta} = -
    K^-_{\alpha\beta} \quad \Rightarrow
    \quad [K_{\alpha\beta}] = 2 K^+_{\alpha\beta},
\end{equation}
which then gives
\begin{equation}\label{SM stress-energy}
    S_{\alpha\beta} = -2 \varepsilon \kappa_5^{-2}
    P^+_{\alpha\beta}.
\end{equation}
Therefore, the stress-energy tensor of conventional matter
\emph{on} the brane is entirely determined by the
\emph{extrinsic} curvature of $\Sigma_0$ evaluated in the $\ell
\rightarrow 0^+$ limit.

The embedding problem takes on a slightly different flavor in the
RS model.  We still want to endow the $\Sigma_0$ 3-brane with
desirable properties, but we must also respect the $\mathbb{Z}_2$
symmetry and the discontinuous nature of the 5-geometry.  We are
helped by the fact that the constraint equations (\ref{4d field
eqns}) are invariant under $K_{\alpha\beta} \rightarrow
-K_{\alpha\beta}$.  This suggests the following algorithm for the
generation of a braneworld model:
\begin{enumerate}
    \item Solve the constraint equations (\ref{4d field eqns}) on
    $\Sigma_0$ for $\Psi^a(y,0) = \Psi_0^+$ such that $\Sigma_0$
    has the desired properties.
    \item Obtain the solution for $\Psi^a(y,\ell)$ for $\ell > 0$ using
    the evolution equations $\di_\ell \Psi^a$ and $\Psi_0^+$ as initial
    data.
    \item Generate another solution of the constraint equations
    by making the switch $K_{\alpha\beta}(y,0) \rightarrow -
    K_{\alpha\beta}(y,0)$ in $\Psi_0^+$.  Call the new solution
    $\Psi_0^-$.
    \item Finally, derive the solution for $\Psi^a(y,\ell)$ for $\ell <
    0$ using $\Psi_0^-$ as initial data.  The resulting solution
    for the bulk geometry will automatically be discontinuous and
    incorporate the $\mathbb{Z}_2$ symmetry about $\Sigma_0$.
\end{enumerate}
This is of course very similar to the standard embedding
procedure already outlined in Sec.~\ref{sec:campbell}, which
allows us to apply the various conclusions of the
Campbell-Magaard theorem to the thin braneworld scenario.  In
particular, we can still arbitrarily choose the induced metric on
$\Sigma_0$ and have enough freedom to consistently solve the
constraint equations. Therefore, \emph{any solution of
$(3+1)$-dimensional general relativity can be realized as a thin
3-brane in the RS scenario}. However, to accomplish this we lose
control of the jump in extrinsic curvature $[K_{\alpha\beta}]$
across $\Sigma_0$, which is related to the stress-energy tensor
of standard model fields living on the brane.  So, if we fix the
intrinsic geometry of the brane then the properties of
conventional matter will be determined dynamically.

We can also consider the inverse of this problem.  Instead of
fixing $h_{\alpha\beta}(y,0)$, we can instead fix
$S_{\alpha\beta}$.  Then, equation (\ref{SM stress-energy}) acts
as $\tfrac{1}{2} n (n+1) = 10$ additional field equations on
$\Sigma_0$ for the elements of $\Psi_0^+$ or $\Psi_0^-$. By a
similar argument as before, this means that we do not have enough
residual freedom to completely choose $h_{\alpha\beta}(y,0)$ or
$K^\pm_{\alpha\beta}$, which means that they are determined
dynamically.  This is a more traditional approach in that the
configuration of conventional matter determines the induced
metric on $\Sigma_0$ (albeit through unconventional field
equations, as described below).  It is interesting to note that
the structure of the constraint equations allows one to either
choose the geometry and solve for the matter, or choose the
matter and solve for the geometry, just like Einstein's
equations.  This similarity means that a generic problem in
general relativity also creeps into the braneworld scenario: the
functional form of $S_{\alpha\beta}$ is not sufficient to
determine the properties of the matter configuration --- one also
needs the metric.  Since $h_{\alpha\beta}$ is determined by the
stress-energy tensor, we cannot have \emph{a priori} knowledge of
the distribution of matter-energy.  As in general relativity, the
way out is to make some sort of \emph{ansatz} for
$h_{\alpha\beta}$ and $S_{\alpha\beta}$ and try to solve for the
geometry and matter simultaneously \cite{Wal84}.

The field equations on the brane are simply given by (\ref{4d
einstein}) with $K_{\alpha\beta}$ evaluated on either side of
$\Sigma_0$.  Usually, equation (\ref{SM stress-energy}) is used
to eliminate $K_{\alpha\beta}^\pm$, which yields the following
expression for the Einstein 4-tensor on $\Sigma_0$:
\begin{eqnarray}\nonumber
    G_{\alpha\beta} & = & \frac{\kappa_5^4}{12} \left[ S
    S_{\alpha\beta} - 3 S_{\alpha\mu} S^\mu{}_\beta +
    \left( \frac{3 S^{\mu\nu} S_{\mu\nu}  - S^2}{2} \right)
    h_{\alpha\beta} \right] \\ & & -\varepsilon E_{\alpha\beta} - \frac{1}{2}
    \lambda h_{\alpha\beta}.\label{thin einstein}
\end{eqnarray}
Since this expression is based on the equations of constraint
(\ref{4d field eqns}), it is entirely equivalent to the STM
expression (\ref{stm einstein}) when $\lambda = 0$.  However, it
is obvious that the two results are written in terms of different
quantities.  To further complicate matters, many workers write
the braneworld field equations in terms of the non-unique
decomposition
\begin{equation}
    S_{\alpha\beta} = \tau_{\alpha\beta} - \tilde{\lambda}
    h_{\alpha\beta},
\end{equation}
so that the final result is in terms of $\tau_{\alpha\beta}$ and
$\tilde{\lambda}$ instead of $S_{\alpha\beta}$.  On the other
hand, STM field equations are often written in a non-covariant
form, where partial derivatives of the induced metric with
respect to $\ell$ appear explicitly instead of $K_{\alpha\beta}$
and $E_{\alpha\beta}$ \cite[for example]{Wes99}. We believe that
this disconnect in language is responsible for the fact that few
workers have realized the substantial amount of overlap between
the two theories; however, we should mention that the
correspondence between ``traditional'' STM and brane world field
equations has been previously verified in a special coordinate
gauge by Ponce de Leon \cite{Pon01}.

Let us now turn our attention to observers in the RSII scenario.
To simplify matters, let us make the 5-dimensional gauge choice
$\Phi = 1$ (our results will of course be independent of this
choice). Then, the $\ell$ equation of motion (\ref{extra EOM})
for observers reduces to
\begin{equation}
    \ddot\ell = \varepsilon
    (K_{\alpha\beta} u^\alpha u^\beta + {\mathfrak{F}}).
\end{equation}
Now by using equation (\ref{K-S relation}), we obtain
\begin{equation}
    K^\pm_{\alpha\beta} u^\alpha
    u^\beta = \mp \tfrac{1}{2} \varepsilon \kappa_5^2 \left[ S_{\alpha\beta} u^\alpha u^\beta - \tfrac{1}{3}
    ( \kappa - \varepsilon \dot\ell^2 )
    S \right].
\end{equation}
We can view this as the zeroth order term in a Taylor series
expansion of $K_{\alpha\beta} u^\alpha u^\beta$ in powers of
$\ell$. In this spirit, the equation of motion can be rewritten
as
\begin{eqnarray}\nonumber
    \ddot{\ell} & = & - \tfrac{1}{2} \mathrm{sgn}(\ell)
    \kappa_5^2 \left[ S_{\alpha\beta} u^\alpha u^\beta - \tfrac{1}{3}
    ( \kappa - \varepsilon \dot\ell^2 )
    S \right] \\ & & + \varepsilon {\mathfrak{F}} +
    O(\ell),\label{another EOM}
\end{eqnarray}
where
\begin{equation}
    \mathrm{sgn}(\ell) =
    \begin{cases}
        +1, & \ell > 0, \\
        -1, & \ell < 0, \\
        \mathrm{undefined}, & \ell = 0,
    \end{cases}
\end{equation}
and we remind the reader that $u^A u_A = u^\alpha u_\alpha +
\varepsilon \xi^2 = \kappa$ (we will assume that $u^A$ is
timelike).  From this formula, it is obvious that freely-falling
observers (${\mathfrak{F}} = 0$) can be confined to a small
region around the brane if
\begin{equation}\label{inequality 1}
    S_{\alpha\beta} u^\alpha u^\beta - \tfrac{1}{3}
    ( \kappa - \varepsilon \dot\ell^2 )
    S > 0.
\end{equation}
Of course, if the quantity on the left is zero or the coefficient
of the $O(\ell)$ term in (\ref{another EOM}) is comparatively
large, we need to look to the sign of the $O(\ell)$ term to decide
if the particle is really confined.  To get at the physical
content of (\ref{inequality 1}), let us make the slow-motion
approximation $\dot\ell^2 \ll 1$. With this assumption, equation
(\ref{inequality 1}) can be rewritten as
\begin{equation}
    \int dy \left\{ T^{(\Sigma)}_{AB} - \tfrac{1}{3}
    \mathrm{Tr}[ T^{(\Sigma)} ] g_{AB} \right\} u^A u^B > 0.
\end{equation}
This is an integrated version of the 5-dimensional strong energy
condition as applied to the brane's stress-energy tensor, which
includes a vacuum energy contribution from the brane's tension.
Its appearance in this context is not particularly surprising;
the Raychaudhuri equation asserts that matter that obeys the
strong energy condition will gravitationally attract test
particles.  Therefore, we have shown that test observers can be
gravitationally bound to a small region around $\Sigma_0$ if the
total matter-energy distribution on the brane obeys the
5-dimensional strong energy condition, and their velocity in the
$\ell$-direction is small.

Finally, we would like to show that the equation of motion
(\ref{another EOM}) has a sensible Newtonian limit.  Let us
demand that all components of the particle's spatial velocity
satisfy $|u^i| \ll 1$ with $i = 1,2,3,4$.  Let us also neglect
the brane's tension and assume that the density $\rho$ of the
confined matter is much larger than any of its principle
pressures.  Under these circumstances we have \cite{Wal84}:
\begin{equation}
    S_{\alpha\beta} u^\alpha u^\beta \approx \rho, \quad
    h^{\alpha\beta} S_{\alpha\beta} \approx \kappa \rho.
\end{equation}
The 5-dimensional coupling constant $\kappa_5^2$ is taken to be
\begin{equation}\label{coupling}
    \kappa_5^2 = \tfrac{3}{2} V_3 G_5,
\end{equation}
where $V_3$ is the dimensionless volume of the unit 3-sphere and
$G_5$ is the 5-dimensional Newton constant.\footnote{The
gravity-matter coupling $\kappa_5^2$ is physically distinguished
from Newton's constant $G_5$ in that the former is the coefficient
of the stress-energy tensor in Einstein's equations while the
latter is the constant that appears in the 5-dimensional
generalization of Newton's law of universal gravitation; i.e., in
the Newtonian limit of the 5-dimensional theory, the gravitation
acceleration around a point mass is $G_5 M/r^3$. This expression
for $\kappa_5^2$ given in equation (\ref{coupling}) is consistent
with the Newtonian force law and the (4+1)-version of Poisson's
equation $\nabla_{(4)}^2 \phi = V_3 G_5 \rho$, where
$\nabla_{(4)}^2$ is the Laplacian operator in Euclidean 4-space.
The fastest way to convince oneself of this is to compare the
Newtonian and general relativistic expressions for the tidal
acceleration between test particles, as in Section 4.3 of Wald
\cite{Wal84}.} With these approximations, we get the following
equation of motion for freely-falling observers:
\begin{equation}
    \ddot{\ell} \approx -\tfrac{1}{2} \mathrm{sgn}(\ell) V_3 G_5
    \rho + O(\ell).
\end{equation}
This is precisely the result that one would obtain from a
Newtonian calculation of the gravitational field close to a
3-dimensional surface layer in a 4-dimensional space using
Gauss's Law:
\begin{equation}
    - \int_{\partial {\mathcal V} } \mathbf{g} \cdot d\mathbf{A} = V_3 G_5
    \int_{\mathcal V} \rho \, d{\mathcal V}.
\end{equation}
Here the integration 4-volume ${\mathcal V}$ is a small
``pill-box'' traversing the brane.  Thus, we have shown that the
full general-relativistic equation of motion in the vicinity of
the brane (\ref{another EOM}) reduces to the 4-dimensional
generalization of a known result from 3-dimensional Newtonian
gravity in the appropriate limit.

In conclusion, we have seen that the Campbell-Magaard theorem
says that it is possible to embed any solution of 4-dimensional
general relativity in the RSII scenario.  However, the price to
be paid is that the matter content of the brane must then be
determined dynamically.  The field equations on the brane were
seen to be similar to those of STM theory.  We also found that
test observers can be gravitationally confined to a small region
around $\Sigma_0$ if the brane's stress-energy tensor obeys the
5-dimensional strong energy condition, and that their
$\ell$-equation of motion reduces to the Newtonian result in the
appropriate limit.  This implies that, as in the STM case, the
requirement that observers be confined to $\Sigma_0$ imposes
restrictions on the embedded spacetime.

\subsection{The thick braneworld scenario}

The last 5-dimensional model of our universe that we want to talk
about is the so-called thick braneworld model \cite{DeW99,Csa00}.
This scenario is essentially a ``smoothed-out'' version of the
RSII picture, where the infinitely sharp domain wall at $\ell =
0$ is replaced with a continuously differentiable 4-dimensional
geometric feature. There are two main motivations for the study
of such an extension of RSII.  First, since there is a natural
minimum length scale in superstring/supergravity theories, the
notion of an infinitely thin geometric defect must be viewed as
an approximation.  Second, one would like to see how these branes
might arise dynamically from solutions of 5-dimensional
supergravity theories, which are by necessity smooth solutions of
some higher-dimensional action involving dilatonic scalar and
other types of fields.  The latter motivation means that the bulk
may contain fields in addition to a non-vanishing vacuum energy
in thick braneworld models.  This means that our previous
formulae cannot be applied to thick braneworlds generated by
scalar fields or other higher-dimensional matter.  For this
reason, we will only consider thick braneworlds where such fields
are absent.  We mention in passing that we expect the results of
this section can be straightforwardly generalized, largely
because we know that the Campbell-Magaard theorem applies to
situations where the higher-dimensional manifold is sourced by
quite general stress-energy tensors \cite{And01b,Dah01b}.

In moving from the RSII scenario to the thick braneworld, we
retain the $\mathbb{Z}_2$ symmetry across the $\Sigma_0$
hypersurface.  In order to satisfy the requirement that the
extrinsic curvature be an odd function of $\ell$, we need to have
that $\Sigma_0$ has $K_{\alpha\beta} = 0$.  This simplifies the
situation enormously.  The Einstein tensor on the brane then
reduces to
\begin{equation}\label{thick einstein}
    G_{\alpha\beta} = - \varepsilon E_{\alpha\beta} -
    \tfrac{1}{2} \lambda h_{\alpha\beta}.
\end{equation}
In contrast to the $K_{\alpha\beta} = 0$ case for STM theory, the
stress-energy of a thick brane embedded in a 5-dimensional
Einstein space does not have vanishing trace unless $\lambda =
0$. Finally, we note that $K_{\alpha\beta} = 0$ allows us to
solve equation (\ref{extra EOM}) with $\ell = 0$: we can always
construct freely-falling test-observer trajectories that lie
within the brane.  Since the extrinsic curvature of $\Sigma_0$ is
fixed by the $\mathbb{Z}_2$ symmetry, the constraint equations
(\ref{4d field eqns}) cannot be solved for arbitrary
$h_{\alpha\beta}$.  So the Campbell-Magaard theorem cannot be
applied to this scenario.

Finally, we note that there is an additional requirement that
must be met by viable thick braneworld models.  Namely, the
graviton ground state must be effectively localized near
$\Sigma_0$.  That is, 5-dimensional metric perturbations must
satisfy a wave equation whose ground state is normalizable and
peaked near the brane.  The issue of normalizability clearly
depends on the global properties of $M$, and is hence not
addressable by the purely local formalism we have introduced thus
far.  At the risk of appearing pedantic, a short comment about
this issue may be in order.  Field equations like those of
Einstein --- or (\ref{thick einstein}) above --- take no notice
of boundary conditions or other considerations such as topology;
and the non-accessibility of such in the cosmological domain has
led Wheeler (following Einstein) to argue that the global
structure of 4-dimensions ought to be closed.  It seems to us
that a parallel argument can be made in the case of the thick
braneworld scenario.  We have no way of knowing the boundary
conditions in 5-dimensions, so in the next section we turn our
attention to a concrete but local example.

\section{An example: a $\mathbb{Z}_2$ symmetric embedding of radiation
dominated cosmology in a 5-dimensional vacuum}\label{sec:example}

In this, our penultimate section, we will consider a concrete
application of the Campbell-Magaard theorem.  Our aim is to take
an empty solution of STM theory and interpret it in terms of the
braneworld scenario.  The induced metric on the brane will be
seen to correspond to standard radiation-dominated cosmology.  We
will also investigate the behaviour of geodesics in the
neighbourhood of the brane and the nature of the induced matter
on hypersurfaces $\Sigma_\ell \ne \Sigma_0$.

We specialize to 5-dimensional manifolds with signature
$(+----)$, which means that the normal to $\Sigma_\ell$ is
spacelike, $\varepsilon = -1$, and timelike geodesics have $u^A
u_A = +1$. Our starting point is a particularly interesting line
element presented in ref.~\cite{Liu01}.  It is given by:
\begin{subequations}
\begin{eqnarray}\label{5D Metric}
    ds^2_{(5)} & = & B^2(t,\ell) dt^2 - A^2(t,\ell) d\sigma_k^2 - d\ell^2,
    \\ \label{A def}
    A^2(t,\ell) & = & [ \mu^2(t) + k ]\ell^2 + 2 \nu(t) \ell + \frac{\nu^2(t)
    + {\mathcal K}}{\mu^2(t) + k}, \\
    B(t,\ell) & = & \frac{1}{\mu(t)} \frac{\di}{\di t} A(t,\ell).
\end{eqnarray}
\end{subequations}
Here $k = 0, \pm 1$ is the curvature index of the 3-geometry,
$d\sigma_k^2$ is the metric on 3-spaces of constant curvature
\begin{subequations}
\begin{eqnarray}
    d\sigma_k^2 & = & d\chi^2 + S_k^2(\chi) (d\theta^2 +
    \sin^2\theta\,d\varphi^2), \\
    S_k(\eta) & \equiv &
    \begin{cases}
        \sin\eta, & k = +1, \\
        \eta, & k = 0, \\
        \sinh\eta, & k = -1,
    \end{cases}
\end{eqnarray}
\end{subequations}
and $\mu(t)$ and $\nu(t)$ are arbitrary functions of time. The
arbitrary constant ${\mathcal K}$ is related to the 5D
Kretschmann scalar via
\begin{equation}
    \hat{R}_{ABCD} \hat{R}^{ABCD} = \frac{72 {\mathcal K}^2}{A^8(t,\ell)}.
\end{equation}
This metric satisfies the 5-dimensional vacuum field equations
\begin{equation}
    \hat{R}_{AB} = 0,
\end{equation}
and is hence a solution of STM theory.  This metric is of
interest because the line element on $\Sigma_\ell$ hypersurfaces
is isometric to standard Friedman-Lema\^{\i}tre-Robertson-Walker
(FLRW) cosmologies, as we will see below.

We now want to use some of the functional arbitrariness in
(\ref{5D Metric}) to obtain a thick braneworld model with
$\mathbb{Z}_2$ symmetry about $\ell = 0$.  Recalling that metrics
with $\mathbb{Z}_2$ symmetry must have components that are even
functions of $\ell$, we see that we should set $\nu(t) = 0$ in
equation (\ref{5D Metric}). If we also make the coordinate
transformation $t \rightarrow \mu = \mu(t)$, we obtain the
following form of the metric:
\begin{subequations}
\begin{eqnarray}\label{5D Metric 2}
    ds^2_{(5)} & = & b^2(\mu,\ell) d\mu^2 - a^2(\mu,\ell) d\sigma_k^2
    - d\ell^2, \\
    a^2(\mu,\ell) & = & ( \mu^2 + k ) \ell^2 + \frac{{\mathcal K}}{\mu^2 + k}, \\
    b(\mu,\ell) & = & \frac{ [( \mu^2 + k )^2 \ell^2 - {\mathcal K} ] }
    {(\mu^2 + k)^{3/2} [( \mu^2 + k )^2 \ell^2 + {\mathcal K} ]^{1/2} } .
\end{eqnarray}
\end{subequations}
This solution is manifestly $\mathbb{Z}_2$ symmetric about $\ell =
0$, implying $K_{\alpha\beta} = 0$ for the $\Sigma_0$
hypersurface.  Notice that to ensure $a(\mu,\ell)$ is real-valued
on the brane at $\ell = 0$, we need to demand
\begin{equation}\label{parameter inequality}
    \frac{\mu^2 + k}{\mathcal{K}} > 0.
\end{equation}
Our field equations (\ref{4d field eqns}) with $\lambda = 0$
predict $G^\alpha{}_\alpha = 0$ on the brane.  This can be
confirmed by direct calculation using the induced metric on
$\Sigma_0$:
\begin{equation}\label{4d metric}
    ds^2_{(\Sigma_0)} = \frac{{\mathcal K}}{\mu^2+k} \left[ \frac{d\mu^2}{(\mu^2 + k)^2} -
    d\sigma_k^2 \right],
\end{equation}
which yields
\begin{equation}
    G^\alpha{}_\beta \Big|_{\Sigma_0} = \frac{(\mu^2 + k)^2}{{\mathcal K}} \left(
    \begin{array}{cccc}
        +3 &    &    &    \\
           & -1 &    &    \\
           &    & -1 &    \\
           &    &    & -1
    \end{array}
    \right).
\end{equation}
Here we have made a choice of 4-dimensional coordinates such that
$e^A_\alpha = \delta^A_\alpha$.  If $G_{\alpha\beta}$ is
interpreted as the stress-energy tensor of a perfect fluid, it has
a radiation-like equation of state $\rho = 3p$.  Also note that
the inequality (\ref{parameter inequality}) implies that the
density and pressure are negative if $(\mu^2 + k) < 0$.  Finally,
if we change 4-dimensional coordinates via
\begin{equation}
    \mu \rightarrow \eta = \eta(\mu) = \int^\mu_{\mu_0} \frac{dx}{x^2 + k},
\end{equation}
and carefully choose $\mu_0$, we get the following line element
on the brane:
\begin{equation}
    ds^2_{(\Sigma_0)} = {\mathcal K} S^2_k(\eta) [ d\eta^2 - d\sigma_k^2
    ].
\end{equation}
This is the standard solution for a radiation-dominated FLRW
cosmology expressed in terms of the conformal time $\eta$
\cite{Pea99}. We have thus obtained a $\mathbb{Z}_2$ symmetric
embedding of a radiation dominated universe in a Ricci-flat
5-dimensional manifold.

Let us now consider the 5-dimensional geodesics of this model in
the vicinity of the brane. The isotropy of the ordinary 3-space
in the model means we can set $\dot{r} = \dot{\theta} =
\dot{\varphi} = 0$ and deal exclusively with comoving
trajectories. The Lagrangian governing such paths is
\begin{equation}
    L = \tfrac{1}{2} \left[ b^2(\mu,\ell) \dot{\mu}^2 - \dot{\ell}^2
    \right].
\end{equation}
We can obtain an equation for $\ddot{\ell}$ by extremizing the
action, to yield:
\begin{equation}\label{ell accn}
    \ddot\ell = -\frac{1}{2} \dot\mu^2 \frac{\di}{\di\ell} b^2(\mu,\ell) =
    \left( \frac{3\dot\mu^2}{\mu^2 + k} \right) \ell + O(\ell^3).
\end{equation}
We see that $\ell = 0$ is an acceptable solution of this equation,
which is reasonable because $\Sigma_0$ has a vanishing extrinsic
curvature.  So, 5-dimensional geodesics can indeed be confined to
the brane.  What is more interesting is the behaviour of geodesics
near the brane.  The coefficient of $\ell$ on the righthand side
of (\ref{ell accn}) is explicitly positive if the density and
pressure of the matter in the embedded brane universe is positive,
so test particles near the brane will experience a force pushing
them away from $\ell = 0$. That is, the 3-brane in this model
represents an unstable equilibrium for observers if the induced
matter on $\Sigma_0$ has reasonable properties.

Finally, we note that we have been primarily concerned with the
$\Sigma_0$ hypersurface.  However, each of the hypersurfaces in
the $\Sigma_\ell$ foliation can be interpreted as a different
4-dimensional universe.  Since the hypersurfaces $\Sigma_\ell \ne
\Sigma_0$ do \emph{not} have $K_{\alpha\beta} = 0$, we expect
that their induced matter does \emph{not} have a radiation-like
equation of state.  To determine the properties of these
universes, we use the induced metric on $\Sigma_\ell$ to
calculate the Einstein 4-tensor, which turns out to be given by:
\begin{subequations}
\begin{eqnarray}
    G^\alpha{}_\beta \Big|_{\Sigma_\ell} & = & \kappa_4^2 \left(
    \begin{array}{cccc}
        + \rho &    &    &    \\
           & -p &    &    \\
           &    & -p &    \\
           &    &    & -p
    \end{array}
    \right), \\
    \kappa_4^2 \rho(\mu,\ell) & \equiv & \frac{3(\mu^2
    +k)}{a^2(\mu,\ell)}, \\
    \kappa_4^2 p(\mu,\ell) & \equiv & \frac{ 2 a(\mu,\ell) + (\mu^2 + k) b(\mu,\ell)
    }{a^2( \mu, \ell) b(\mu,\ell)}.
\end{eqnarray}
\end{subequations}
From these expressions for the density and
pressure of the induced matter on $\Sigma_\ell$, we can derive
the following expression for the so called quintessence
parameter:
\begin{equation}
    \gamma(\mu,\ell) = \frac{ p(\mu,\ell) }{ \rho(\mu,\ell) } =
    \frac{1}{3} \left[ \frac{ {\mathcal K} + 3 \ell^2 ( \mu^2 + k )^2 } {
    {\mathcal K} - \ell^2 ( \mu^2 + k )^2 } \right].
\end{equation}
For $\ell = 0$, we recover our previous result $\gamma = 1/3$ for
all $\mu$. For $\ell \ne 0$, we obtain $\gamma \rightarrow -1$ as
$\mu \rightarrow \infty$.  Hence, the universes located at $\ell
\ne 0$ approach the vacuum-dominated FLRW solution (i.e., $\rho =
-p$) for late times.  These results are very plausible from the
physical perspective.

What we have done immediately above may be summarized with a view
to future work.  We have used the arbitrariness in a known
solution of STM theory to generate a $\mathbb{Z}_2$ symmetric
embedding of FLRW radiation-dominated cosmologies in a
5-dimensional Ricci-flat manifold.  The $\Sigma_0$ hypersurface is
an example of how a 3-brane with zero extrinsic curvature has
$G^\alpha{}_\alpha = 0$ when $\lambda = 0$.  We have also
presented a concrete realization of the confinement of
5-dimensional test observers to a 3-brane. However, the
equilibrium position of observers on $\Sigma_0$ was shown to be
unstable if the density and pressure of the induced matter on the
brane is positive. The 4-dimensional spacetimes corresponding to
the $\Sigma_\ell$ hypersurfaces other than $\Sigma_0$ were seen to
approach the de Sitter FLRW universe for late times.  These
results are intriguing, but preliminary. We note that the
components of $g_{AB}$ in this model are non-separable functions
of $\mu$ and $\ell$, so a complete analysis of the tensor and
scalar waves admitted by this model will be non-trivial.

\section{Summary}\label{sec:conclusions}

In this paper we have derived field equations (\ref{4d field
eqns}) for three spin-2 fields $\{ h_{\alpha\beta},
K_{\alpha\beta}, E_{\alpha\beta} \}$ living on an $n$-dimensional
hypersurface $\Sigma_0$ embedded in an $(n+1)$-dimensional
Einstein space $M$. We have used these equations to give a
heuristic proof of a generalized Campbell-Magaard theorem, which
states that it is possible to embed any $n$-dimensional
psuedo-Riemannian manifold in an $(n+1)$-dimensional space with
or without a cosmological constant. We also demonstrated that
instead of embedding $\Sigma_0$ in $M$ with an arbitrary metric
$h_{\alpha\beta}$, we can instead embed it with arbitrary
extrinsic curvature $K_{\alpha\beta}$.

These results were then applied to three different 5-dimensional
models of the universe. In STM theory, we found that the theorem
allowed us to realize any solution of general relativity as a
4-surface in a 5-dimensional vacuum spacetime.  However, by
examining a $(4+1)$-splitting of the equation of motion of test
particles, we found that observers will not be confined to
$\Sigma_0$ unless it has vanishing extrinsic curvature.  If the
latter condition is imposed, then the induced matter on
$\Sigma_0$ has a radiation-like equation of state. In the RSII
thin braneworld scenario, we found that one could also embed any
solution of 4-dimensional relativity on a 3-brane. However, in so
doing one loses control of the stress-energy tensor of standard
model fields living on $\Sigma_0$.  We showed that test observers
in this model can be confined to a small region around $\Sigma_0$
if the total brane stress-energy tensor obeys the 5-dimensional
strong energy condition. Finally, we found that the
$\mathbb{Z}_2$ symmetry in the thick braneworld model
\emph{requires} that $K_{\alpha\beta} = 0$ on the brane. This
implies a restrictive 4-geometry, and we cannot embed arbitrary
spacetimes if the bulk contains only vacuum energy.  Also, test
observers are naturally confined to the brane in this scenario.

We considered for illustrative purposes a class of solutions from
STM theory.  This class included enough arbitrariness for us to
impose the $\mathbb{Z}_2$ symmetry about the $\ell = 0$
hypersurface. We found that the induced matter on the brane had a
radiation-like equation of state, in agreement with previous
results.  Using 4-dimensional coordinate transformations, we
demonstrated that the geometry on $\Sigma_0$ exactly matched the
standard radiation-dominated FLRW models.  We also found that
$\ell = 0$ was an unstable equilibrium for test observers if the
density and pressure in those models is positive. We also looked
at the properties of the universes on $\ell \ne 0$ hypersurfaces
and found that they are asymptotically the same as de Sitter FLRW
cosmologies.

One of the main themes that has emerged from our considerations
is the mathematical similarity between the STM and braneworld
scenarios.  Despite the fact that they have very different
motivations, we have found that the field equations of each
theory are definitely related.  The thin braneworld expression
for $G_{\alpha\beta}$ (\ref{thin einstein}) reduces to the STM
formula (\ref{stm einstein}) in the $\lambda \rightarrow 0$ limit
and with an appropriate choice of gauge. The $K_{\alpha\beta} =
0$ case of STM theory matches the $\lambda = 0$ case (\ref{thick
einstein}) in the thick braneworld scenario we considered.  Also,
the thin braneworld field equations reduce to the thick
braneworld case when $K_{\alpha\beta} = 0$.  The reason for these
correspondences, which have been noted before \cite{Pon01}, is
seen to be something rather simple: If we wish to embed our
4-dimensional matter-filled world in an empty 5-dimensional
universe, the constraint equations (\ref{4d field eqns}) have to
be obeyed.  A technical corollary of this is that the numerous
known solutions of STM theory \cite{Wes99} can be reinterpreted
as brane models.

\begin{acknowledgments}
We would like to thank W.~Israel for comments on the
Campbell-Magaard theorem and the motion of observers around
surface layers, an anonymous referee for pointing an error in the
original manuscript, and NSERC for financial support.
\end{acknowledgments}

\appendix

\section{Covariant splitting of test particle equations of
motion}\label{sec:EOM}

In ref.~\cite{Sea02}, a covariant $(4+1)$-splitting of the
5-dimensional geodesic equation was performed for the case $n^A
n_A = -1$.  This resulted in two equations of motion, one for
motion parallel to $\Sigma_\ell$ and another for motion
orthogonal to $\Sigma_\ell$. We propose to generalize those
results to the current situation where $n^A n_A = \varepsilon$,
and to include the possibility that observers' trajectories are
subject to the influence of some non-gravitational force per unit
mass, denoted by $f^A$.  The relevant 5D equations of motion are
given by:
\begin{subequations}
\begin{eqnarray}
    u^B \nabla_B u^A & = & f^A, \\ u_A u^A & = & \kappa \equiv
    +1,0,-1, \\
    u^A & \equiv & dx^A / d\lambda \equiv \dot{x}^A.
\end{eqnarray}
\end{subequations}
Here $\lambda$ is the 5-dimensional affine parameter, and an
overdot indicates $d/d\lambda = u^A \di_A$.  Generalizing the
calculations found in ref.~\cite{Sea02}, the 5-dimensional
equations of motion can be written as
\begin{subequations}\label{split equations}
\begin{eqnarray}\label{4D part}
    u^{\beta}{}_{;\alpha} u^\alpha  & = & -\varepsilon \xi (K^{\alpha\beta}
    u_\alpha + e^\beta_B n^A \nabla_A u^B) + f^\beta, \\ \label{extra part}
    \dot{\xi} & = & K_{\alpha\beta} u^\alpha u^\beta + \varepsilon \xi n^A
    u^B \nabla_A n_B + {\mathfrak{F}}, \\ \label{split norm}
    \kappa & = & h_{\alpha\beta} u^\alpha u^\beta + \varepsilon \xi^2.
\end{eqnarray}
\end{subequations}
These involve the definitions
\begin{subequations}
\begin{eqnarray}
    u^\alpha & \equiv & e^\alpha_A u^A, \quad
    \xi \equiv n_A u^A, \\
    f^\alpha & \equiv & e^\alpha_A f^A, \quad
    {\mathfrak{F}} \equiv n_A f^A.
\end{eqnarray}
\end{subequations}
It was also demonstrated in ref.~\cite{Sea02} that
\begin{equation}
    u^\alpha = \dot{y}^\alpha + N^\alpha
    \dot{\ell}, \quad \xi = \varepsilon \Phi
    \dot{\ell},
\end{equation}
where $\Phi$ and $N^\alpha$ are the lapse and shift introduced in
Section \ref{sec:geometry}.  Consider the last relation and the
the identities
\begin{subequations}
\begin{eqnarray}
    (\ln \Phi)_{;\beta} & = & -\varepsilon e_\beta^B n^A
    \nabla_A n_B, \\ \dot{\Phi} & = &
    u^\alpha \Phi_{;\alpha} + \varepsilon \xi n^A \nabla_A
    \Phi.
\end{eqnarray}
\end{subequations}
Then equation (\ref{extra part}) can be rewritten in the mixed
but instructive form
\begin{equation}\label{extra EOM}
    \ddot{\ell} = \frac{\varepsilon}{\Phi} (
    K_{\alpha\beta} u^\alpha u^\beta + {\mathfrak{F}}) - \dot\ell \left[ 2 u^\beta
    (\ln\Phi)_{;\beta} + \dot\ell n^A \nabla_A \Phi
    \right].
\end{equation}
In this paper, we will be primarily concerned with this
expression, which governs motion perpendicular to $\Sigma_\ell$.

We conclude by saying a few words about the $\kappa$ parameter:
Our formalism for dealing with 5-dimensional geodesics can be
applied to timelike, spacelike or null paths. For each of these
cases, a choice of metric signature must be made before $\kappa$
can be specified. For example, if the 5-dimensional metric
signature is $(+---\pm)$, then timelike paths have $u^A u_A = +1$
or $\kappa = +1$.

%\bibliography{text}

\end{document}